 \documentclass{aastex}

\begin{document}

\title{Optimal detection of sources \\
    on a homogeneous \& isotropic background}

\author{J. L. Sanz$^{1}$, D. Herranz$^{1,2}$ 
and E. Mart\'\i nez-G\'onzalez$^{1}$}

\affil{ (1) Instituto de F\'\i sica de Cantabria,  
             Fac. de Ciencias, Av. de los Castros s/n, \\
             39005-Santander, Spain \\
 (2) Departamento de F\'\i sica Moderna, 
             Universidad de Cantabria,	     
             39005-Santander, Spain}

\begin{abstract}
This paper  introduces the use of pseudo-filters that optimize
the detection/extraction of sources on a background. We assume as a first 
approach that such sources
are described by a spherical (central) profile and that the background is 
represented by a homogeneous
\& isotropic random field. We make an n-dimensional 
treatment making emphasis
in astrophysical applications for spectra, images and volumes, 
for the cases of exponential 
and Gaussian source profiles and scale-free power spectra to represent the background.
\end{abstract}

\keywords{methods: analytical --- methods: data analysis}

\section{Introduction}
One of the challenges in the analysis of 1D spectra, 2D images or 3D volumes in
Astrophysics and Cosmology is to overcome the problem of separating a localized signal 
from a background. In particular, we are interested in localized sources with central 
symmetry and a background that we shall assume with the properties of homogeneity and 
isotropy and it will be characterized by a power spectrum. 
Typical cases in the 1D case include:
a) the spectra of QSOs, how to detect/extract absorption lines from a noisy 
spectrum once a profile is assumed, b) time series analysis where a localized emission 
is superposed on a background, c) point and extended sources to be detected in time-ordered 
data where the scanning strategy (for satellites like Planck) is affected by the well-known
 $1/f$ noise. In the 2D case, we mention as typical cases: a) cleaning of images 
to detect typical astrophysical sources on a white noise background, b) the 
detection/extraction of point sources and extended sources (clusters of galaxies) 
on microwave/IR maps where the background is dominated by white noise or intrinsic CMB 
signal or galactic emission. In the 3D case, we remark as an example: a) the detection 
of clusters and large-scale structure in 3D catalogs.

The classical treatment to remove the background has been {\it filtering}. Low and 
high-pass filters reduced the contribution from the high and low frequencies present in 
the spectrum or image. In general, this process is extremely inefficient in dealing with 
very localized sources. The reason for that is that a very localized source can be 
decomposed in Fourier space but many waves are needed (infinite for a delta distribution!), 
so if low/high-pass filters are applied then at the end many artefacts (rings in 2D images) 
usually appear surrounding the sources.

   A very important application of these principles is the detection
of sources in two-dimensional astronomical images. 
Several packages are commonly used for this task, such as
DAOfind (Stetson 1992), used to find stellar 
objects in astronomical images, and 
SExtractor (Bertin \& Arnouts 1996).
When trying to
detect sources, two different problems have to be solved: first,
it is necessary to take account for the background variations across
the image. In addition, if instrumental noise (i. e. white noise)
appears, it should be removed as far as possible in order to
increase the SNR ratio. 
SExtractor estimates the local background on a grid of size set by
the user and then approximates it by a low-order polynomial.
Once such a background is defined the detection of
sources at a certain level can be established
through sets of connected pixels above certain threshold. 
DAOfind implicitly assumes that the background is smooth and its
characteristic scale is very much larger than the scale of the
stars. But in the case where the characteristic scale of 
variation of the background is approximately the scale of the structures 
the previous schemes obviously fail. An incorrect estimation of the
background leads to a biased estimation of the amplitude
of the sources.
An example is a typical image of 
the cosmic microwave background radiation (CMB) at a 
resolution of several arcmin. If the intrinsic signal is due
to a cold dark matter model (i. e. the characteristic scale of
the background is of the order of $\sim 10$ arcmin), then
the separation of point sources with the same
characteristic scale becomes a very difficult task.

To deal with the instrumental noise the traditional procedure 
is filtering (e. g. Gaussian window). DAOfind filters with 
an elliptical Gaussian that mimics the stellar psf
and then it performs the detection looking for peaks 
above certain threshold. SExtractor
includes the possibility of filtering with several kind of
filters (Top Hat, Gaussian, Mexican Hat and even self-made
filters for every particular situation). An obvious advantage of
this procedures (background estimation plus filtering) is that
no a priori information on the structures is needed. A serious
drawback is that the choice of the filter will have a great
influence on the final result. The choice of filter
depends on many factors, including in most cases personal 
preferences. 

  In this context, it is necessary to find a systematic
way to determine the optimal filter for every case.

Other methods have been used to separate different components given several microwave maps: 
Wiener filtering (WF) and maximum entropy methods (MEM). Regarding point sources, WF assumes 
a mean spectral flux dependence together with other assumptions for the other components 
(Tegmark \& Efstathiou, 1996; Bouchet et al. 1999) whereas MEM assumes that the point sources are 
distributed like noise (Hobson et al. 1999). These methods are powerful regarding extended 
sources like clusters 
of galaxies because they use the concrete spectral dependence for the Sunyaev-Zeldovich
effect. It is clear that the unknown  spectral dependence for point
sources remark the inefficiency of the previous methods.

A possible solution to overcome this problem came with the usage of {\it wavelets}. These are 
localized bases that allow a representation of a local object due, in general, to their 
basic properties: spatial and frequency localization (as opposed to the trigonometric functions 
appearing in Fourier decomposition). We remark at this point the success of such a technique 
dealing with simulated microwave maps: the "Mexican Hat" wavelet can be used in a nice way 
(no extra assumptions about the background) to detect/extract point sources and create a 
simulated catalog (Cay\'on et al. 2000, Vielva et al. 2000). Two advantages emerge: 
on the one hand, one localizes the structures associated to the maxima of the wavelet coefficients
and, what is more remarkable, we gain in the detection (as compared to real space) due to the
amplification effect because at the scale of the source the background is not contributing to the 
dispersion. One relevant question concerns the possibility to find {\it optimal} filters.
Tegmark \& Oliveira-Costa (1998) introduced a filter that minimizes the variance in the map.
With this method one can identify a big number of point sources in CMB maps. However, they failed 
to introduce the appropriate constraints in the minimization problem, i. e.
the fact that we have a maximum at the source position at the scale defined by the source in
order not to have spurious identifications.

Thus, this type of analysis lead us to the following questions: is there an optimal filter 
(or better pseudo-filter) given the source profile and the power spectrum of the background?,
Is the "Mexican Hat" wavelet the optimal pseudo-filter dealing with point sources? 
In order to answer these questions, we will assume that the sources can be approximated by 
 structures with central symmetry 
given by a profile $\tau (x)$, $x\equiv |\vec{x}|$ with a 
characteristic scale (e.g. a single
maximum at its center and rapid decay at large distances). 
If the nD image contains different types of sources, a number of  pseudo-filters
adapted to each profile 
must be used to detect them.
A possible 
generalization to include more general profiles is under study. The background will be modelled 
by a homogeneous and isotropic random field given in terms of the power spectrum $P(q)$, 
$q\equiv |\vec{q}|$. In particular, we shall explore a scale-free spectrum 
$P(q)\propto q^{-\gamma}$ that includes the cases of white noise ($\gamma = 0$), $1/f$ noise 
($\gamma = 1$), etc. Moreover, any spectrum of physical interest often can be locally 
approximated by a power-law. If the characteristics of the noise are not known a priori 
it  can be always estimated directly from the nD image. 
We consider the n-dimensional case and make special emphasis on the
analysis of spectra ($n = 1$), 2D images ($n = 2$) and 3D volumes ($n = 3$).
In all the calculations we assume that the overlapping by nearby sources is negligible and also 
that their contribution to the total power spectrum is also negligible. All of this is a very 
good approximation at least above a certain flux level.

An analytical approach to get the optimal pseudo-filter is presented in section~\ref{optfilter}. 
Section~\ref{realspace} deals 
with properties of the optimal pseudo-filters on real space. Section~\ref{detection} 
introduces the concepts of
detection level and gain. Sections~\ref{gaussian} and~\ref{exponential} are 
dedicated to the important cases of sources with
profiles described by a Gaussian and an exponential, respectively. An example of the performance
of optimal pseudo-filters  applied to simulated one-dimensional data is presented in
section~\ref{simulations}.
Section~\ref{extraction} deals with the 
extraction of the sources. Conclusions are summarized in section~\ref{conclusions}.

\section{Optimal pseudo-filter} \label{optfilter}

Let us consider an n-dimensional ($n = 1, 2, 3$) image with data values
defined by

\begin{equation}
y(\vec{x}) = s(x) + n(\vec{x}),
\end{equation}

\noindent where $x$ is the spatial coordinate (in the 1D case $x$ can be also time,
when we are dealing with time-ordered data sets)
 and $s(x), x\equiv |\vec{x}|$, represents a source with central 
symmetry placed at the origin with a characteristic scale (e. g. a single maximum at its center
and rapid decay at large distances) and $n(\vec{x})$ a homogeneous \& isotropic
background (random field) with mean value $<n(\vec{x})> = 0$ and characterized 
by the power spectrum $P(q), q\equiv |\vec{q}|$ 
(this can represent instrumental noise and/or 
a real background),

\begin{equation}
<n(\vec{q})n^*(\vec{q^{\prime}})> = P(q)\delta^n(\vec{q} - \vec{q^{\prime}}),
\end{equation}

\noindent where $n(\vec{q})$ is the nD Fourier transform ($n(\vec{q}) = (2\pi)^{-n/2}
\int d\vec{x} \, e^{-i\vec{q}\vec{x}} n(\vec{x})$), symbol $n^*$ represents the complex 
conjugate of $n$, $\vec{q}$ is the wave vector
and $\delta^n$ is the
nD Dirac distribution.

Let us introduce a spherical (centrally symmetric) 
pseudo-filter, $\Psi (\vec{x}; \vec{b}, R)$, 
dependent on $n + 1$ parameters . 

\begin{equation}
\Psi (\vec{x}; R, \vec{b}) = \frac{1}{R^n}\psi (\frac{|\vec{x} - \vec{b}|}{R}),
\end{equation}

\noindent where $\vec{b}$ defines a translation whereas $R$ defines
a scaling. Then, we define the pseudo-filtered field $w(R, \vec{b})$

\begin{equation}
w(R, \vec{b}) = \int d\vec{x}\,y(\vec{x})\Psi (\vec{x}; \vec{b}, R).
\end{equation}

\noindent We do not assume a priori the positiveness of $\Psi$. 
The previous convolution can be written as a product in Fourier space, in the form

\begin{equation}
w(R, \vec{b}) = \int d\vec{q}\,{e^{-i\vec{q}\vec{b}}}y(\vec{q})\psi (Rq).
\end{equation}

\noindent where $y(\vec{q})$ and $\psi(q)$ are the Fourier transforms of
$y(\vec{x})$ and $\psi(\vec{x})$, respectively. Because of the central symmetry
assumed for the pseudo-filter,
$\psi(q)$ depends only on the modulus of $\vec{q}$. 
A simple calculation -taking into account eqs. (1) and (2)- gives 
the average at the origin $\vec{b} = 0$, 
$<w(R, \vec{0})>$, and the variance,
$\sigma^2_w(R) = <w^2(R, \vec{b})> - <w(R, \vec{b})>^2$, of the pseudo-filtered
field

\begin{equation}
<w(R, \vec{0})> = \alpha \int dq\,q^{n - 1}\,s(q)\psi (Rq),\, 
\sigma^2_w(R) = \alpha \int dq\,q^{n - 1}\,P(q){\psi}^2(Rq).
\end{equation}

\noindent where $q = |\vec{q}|$, $\alpha = 2, 2\pi , 4\pi $ for $n = 1, 2, 3$, 
respectively, (for n-dimensions $\alpha = 2{\pi}^{n/2}{\Gamma}^{-1}(n/2)$) 
and the limits in the integrals go from $0$ to $\infty $.

Now, we are going to express the conditions in order to obtain an optimal
pseudo-filter for the detection of the source $s(x)$ at the origin. One basic idea is to
find a pseudo-filter such that when the original image is filtered with a scale $R_o \approx R_s$
-being $R_s$ the characteristic scale of the source- one obtains the maximum {\it detection level}
${\mathcal D}_w$
 
\begin{equation}
{\mathcal D}_w \equiv \frac{<w(R, \vec{0})>}{\sigma_w(R)}. 
\end{equation}

\noindent Taking into account the fact that the source is characterized by a single scale 
$R_s$, other basic idea is to generate a filter giving the maximum contribution at the 
center of the source at a filtering scale $R_o \approx R_s$. Finally, we would like to 
estimate directly the amplitude of the source by the previous number. Therefore, taking into 
account these basic ideas we will introduce from the mathematical point of view the 
optimal pseudo-filters.

By definition a pseudo-filter will be called {\it optimal} if the following conditions 
are satisfied:

i) there exists a scale $R_o$ such that $<w(R, \vec{0})>$ has a maximum at that
scale, ii) $<w(R_o, \vec{0})> = s(0) \equiv A$, i. e. $w(R, \vec{0})$ is an 
unbiased estimator of the amplitude of the source and iii) the variance of
$w(R, \vec{b})$ has a minimum at the scale $R_o$, i. e. we have an efficient
estimator. As a by-product, the ratio given by eq. (7) will be 
maximum. We remark that no other information about the source profile
is assumed, so ``optimal'' must be understood in the previous 
sense.

 By introducing the profile $\tau (x)$ of the source, $s(x) = A\tau (x)$, the 
condition ii) and the equation (6) give the constraint

\begin{equation}
\int dq\,q^{n - 1}\,\tau (q)\psi (R_oq) = \frac{1}{\alpha},
\end{equation}

\noindent whereas the condition i) gives the constraint

\begin{equation}
\int dq\,q^{n - 1}\,\tau (q)\psi (R_oq) [n + \frac{d\,ln\tau}{d\,lnq}] = 0.
\end{equation}

So, the problem is reduced to the functional minimization 
(with respect to
$\psi$) of $\sigma^2_w(R)$ given by
equation (6) with the constrains given by equations (8) and (9). 
This minimization incorporate these constraints through a couple
of Lagrangian multipliers.
The
solution (optimal pseudo-filter) is found to be

\begin{equation}
\tilde{\psi}(q)\equiv \psi (R_oq) = \frac{1}{\alpha}\frac{\tau (q)}{P(q)}\frac{1}{\Delta}
[nb + c - (na + b) \frac{d\,ln\tau}{d\,lnq}], \ \ \ 
\Delta = ac - b^2,
\end{equation}

\begin{equation}
a\equiv \int dq\,q^{n - 1}\,\frac{{\tau}^2}{P} ,\ \ \ 
b\equiv \int dq\,q^{n - 1}\,\frac{\tau}{P} \frac{d\,\tau}{d\,lnq},\ \ \ 
c\equiv \int dq\,q^{n - 1}\,\frac{1}{P}{[\frac{d\,\tau}{d\,lnq}]}^2.
\end{equation}

Therefore, we have obtained analytically the functional form of the pseudo-filter (its 
shape and characteristic scale are associated to the source profile and power spectrum).
It is clear that assuming an adimensional dependence $\tau (x/R_s)$, where $R_s$ is the 
characteristic scale of the source, then such scale will appear explicitly in the form 
$\tilde{\psi}(qR_s)$. Obviously, we assume all the differentiable and regularity conditions at 
$q\rightarrow 0, \infty $ for $\tau $ and $P(q)$ in order to have finite expressions for
$a, b, c$.
Generically, $\psi $ is a pseudo-filter, i. e. it is not positive (filter). 
Let us remark that if we assume the behavior $\frac{\tau (q)}{P(q)}\rightarrow 0$,
$\frac{1}{P}\frac{d\,\tau}{d\,lnq}\rightarrow 0$ for $q\rightarrow 0$
then $\psi (q)\rightarrow 0$ and $\Psi$ is a "compensated" filter, i. e. 
$\int d\vec{x}\,\Psi = 0$. Strictly speaking there is another condition to be 
satisfied to get the reconstruction of the image and thus to have a wavelet: 
$\int dqq^{-1}\psi^2(q) < \infty $ (the admissibility condition).

Taking into account eq. (5) the amplitude will be estimated as

\begin{equation}
A = w(R_o, \vec{0}) = \int d\vec{q}\,y(\vec{q})\,\tilde{\psi}(q),
\end{equation}

\noindent where $\tilde{\psi}$ is given by eq.(10).

\vskip 1cm

\section {Optimal pseudo-filter on real space} 	\label{realspace}    

The equation (10) can be written on real space as follows

\begin{equation}
\Psi(\vec{x}; R_o, \vec{0}) = \frac{1}{R_o^n}\psi(\frac{|\vec{x}|}{R_o}) = 
\frac{1}{\alpha \Delta }[(nb + c)F(x) - (na + b)G(x)],
\end{equation}

\noindent where $F$ and $G$ are the inverse Fourier transform of $\frac{\tau}{P}$ and
$\frac{1}{P}\frac{d\tau}{dlnq}$, respectively. 

   For a flat background, i. e. $P = constant$, and assuming the behaviour 
$q^{\frac{1}{2}(n + 1)}\tau \rightarrow 0 $ when $q\rightarrow \infty$, one obtains

\begin{equation}
b = -\frac{n}{2}a,\ \ \ 
F = \frac{\tau (x)}{P},\ \ \ G(x) = -\frac{1}{P}[n\tau + \frac{d\tau }{dlnx}].
\end{equation}

If we also assume a Gaussian profile, i. e. $\tau = e^{-\frac{x^2}{2{\theta}^2}}$, one 
finds the pseudo-filter

\begin{equation}
\Psi(\vec{x}; \theta , \vec{0}) = \frac{1}{{\pi}^{n/2}{\theta}^n}
e^{-\frac{x^2}{2{\theta}^2}}[1 + \frac{n}{2} - {(\frac{x}{\theta})}^2],
\end{equation}

\noindent that is a useful formula to be used for the detection of nD-Gaussian structures
on nD-images (e.g. spectra, 2D images or 3D volumes) on a flat background.

On the other hand, if one assumes a Gaussian profile but a non-flat spectrum one can
easily find

\begin{equation}
\Psi(\vec{x}; \theta, \vec{0}) = \frac{1}{\alpha \Delta }
[(nb + c)F(x) - (na + b) {\theta}^2{\nabla}^2F(x)],
\end{equation}

\noindent being $F(x)$ the Fourier transform of $\frac{\tau}{P}$.

\vskip 1cm

\section{Detection level, gain and reliability} \label{detection}

Taking into account the previous expression (10), one can 
calculate the detection level (see equation(7))

\begin{equation}
{\mathcal D}_w  = A{[\frac{\alpha \Delta}{n^2a + 2nb + c}]}^{1/2}.
\end{equation}

On the other hand, we can calculate the dispersion of the field $n(\vec{x})$ 

\begin{equation}
{\sigma}^2_b = \alpha \int dq\,q^{n - 1}P(q),
\end{equation}

\noindent that allows to define a detection level on real space as

\begin{equation}
{\mathcal D} = \frac{A}{\sigma_b}.
\end{equation}

\noindent The {\it gain} going from real space to pseudo-filter space is defined by

\begin{equation}
g \equiv \frac{{\mathcal D}_w }{{\mathcal D}} = \frac{\sigma_b}{\sigma_w(R_o)}.
\end{equation}

\noindent If the background has a characteristic scale different from the scale of the structures
(sources) to be detected, it is obvious that $\sigma_w < \sigma_b$, so that we have a real gain going
from real space to pseudo-filter space.

The identification of sources as peaks above a high threshold (e. g. $3\sigma_w $) in pseudo-filter space 
gives a low probability of false detections (reliability) because if the background has a
characteristic scale different from the sources then everything detected with our method is real, but if
both scales are comparable one can give an estimate based on the fluctuations of the background. For 
instance, in the case of a Gaussian background,
false detections above $3\sigma_w$, due to the Gaussian background, appear with a probabilty
$\simeq 1.5\times 10^{-3}$ based on the formula 

\begin{equation}
Pr( w > A ) = \frac{1}{2}erfc(\frac{A}{2^{1/2}\sigma_w }).
\end{equation}

\noindent To select the false detections from the real ones one can study the pseudo-filter profile nearby
any real source (see the last paragraph of section 8).

Regarding the completeness (i. e. how many real sources we miss with our 
method), this is a complicated topic because the background
can slightly modify the location of the peaks and their amplitude. We will address this problem via
numerical simulations (see section 7).

\section {Gaussian source on a background} \label{gaussian}

In many physical applications the standard response of the instruments can be 
approximated by a Gaussian function. In particular, the Point Spread Function (PSF)
for many telescopes is of Gaussian type. Other more specific applications are related to
the Cosmic Microwave Background (CMB), where the antennas used are well approximated by 
a Gaussian. Dealing with absorption systems associated to QSOs, if the absorption line	
is not saturated and is dominated by thermal motions, then the line is usually approximated 
by an inverted Gaussian inserted in a continuum plus noise.
    
Let us assume that the source and the background can be represented by

\begin{equation}
s(x) = A\,e^{-\frac{x^2}{2{\theta}^2}},\ \ \ P = Dq^{-\gamma}.
\end{equation}

The structure to be detected could have an intrinsic Gaussian profile or it could be 
a point source in n-dimensions observed with an instrument that can be modelled
through a Gaussian pattern with a beam size $\theta$. The background can be
described by a scale-free spectrum. In this case: 
$\tau (q) = {\theta}^ne^{-\frac{1}{2}{(q\theta )}^2}$,
and equations (11) give

\begin{equation}
a = \frac{{\theta}^{n - \gamma}}{2D}\Gamma (m),\ \ \ b = - ma,\ \ \ c = m(1 + m)a, \ \ \ 
m\equiv \frac{n + \gamma}{2}
\end{equation}

\noindent and the pseudo-filter is

\begin{equation}
\tilde{\psi} (q) = \frac{1}{\alpha \Gamma (m)}
{(q\theta)}^{\gamma}e^{-\frac{1}{2}{(q\theta )}^2}[2 + \gamma - n +
\frac{n - \gamma}{m}{(q\theta)}^2].\ \ \ 
\end{equation}

Taking into account the $q$ behaviour in this formula, one obtains a compensated filter 
(i. e. $\psi (q=0) = 0$) if  $\gamma > 0$ or $\gamma = n - 2$.
In figure~\ref{fig1} 
 appear the optimal pseudo-filters for the 1D, 2D and 3D cases, 
respectively, and scale-free power spectrum with indexes $\gamma = 0, 1, 2, 3$.
There is a degeneration in the case $n=3$, where
the $\gamma=1$ line overlaps with the $\gamma=3$ line, and in the case
$n=2$, where the $\gamma=0$ line overlaps with the $\gamma=2$ one.
This degeneration can be deduced directly from equation (23).

On the other hand, the detection level in pseudo-filter space is 
given by equation (18)

\begin{equation}
{\mathcal D}_w= A{[\frac{2\alpha}{D}\frac{\Gamma (1 + m)}
{4m + {(n -\gamma )}^2}]}^{1/2}{\theta}^{\frac{n - \gamma }{2}}. 
\end{equation}

Finally, it is interesting to remark that the cases $\gamma = n$ and $\gamma = n - 2$ give 
the same pseudo-filter

\begin{equation}	    
\tilde{\psi} (q) = \frac{2}{\alpha \Gamma (n)}{(q\theta)}^ne^{-\frac{x^2}{2{\theta}^2}}.
\end{equation}

Therefore, in the cases $n=2, \gamma=0,2$ the Mexican Hat is found to be the optimal
pseudo-filter. This justify the use of this wavelet to detect point sources in
CMB maps (Cay\'on 2000, Vielva 2000).

a) Gaussian source and white noise:	    
	    
This subcase corresponds to $P = constant$ or $\gamma = 0$, and the pseudo-filter is

\begin{equation}
\tilde{\psi} (q) = \frac{2}{\alpha \Gamma (\frac{n}{2})}
e^{-\frac{1}{2}{(q\theta )}^2}[1 - \frac{n}{2} + {(q\theta)}^2],\ \ \ 
\end{equation}

\noindent that gives a pseudo-filter for the analysis in the different dimensions 
except for $n = 2$ that gives the Mexican Hat wavelet 
($\psi = \frac{1}{\pi}e^{-\frac{1}{2}{(q\theta)}^2}{(q\theta)}^2$).

 The detection level is given by equation (24): 
 
\begin{equation}
{\mathcal D}_w = A{[\frac {\alpha \Gamma (\frac{n}{2})}
{2D(1 + \frac{n}{2})}]}^{1/2}{\theta}^{n/2}.  
\end{equation}

 We have calculated the contribution of sources to the 
 power spectrum in order to estimate their influence in calculating the pseudo-filter. 
 We arrive to the conclusion that if the signal/noise ratio (i. e. dispersion associated 
 to the sources over dispersion associated to the background) is 
 $\frac{{\sigma}_s}{{\sigma}_b} < 0.6 (\frac{l_p}{\theta})$, being $l_p$ the pixel scale
and $\theta$ the width of the source, then
 the extra contribution to the pseudo-filter coefficients $a, b, c$ is less than a $10\%$.

b) Gaussian source and $1/f$ noise:
	    
Let us assume a source with a Gaussian profile and a background represented by
$1/f$ noise, i.e. $P = Dq^{-1}$ or $\gamma = 1$.
In this case: $\tau (q) = {\theta}^ne^{-\frac{1}{2}{(q\theta)}^2}$,
and equation (22) gives the pseudo-filter

\begin{equation}
\tilde{\psi} (q) = \frac{1}{\alpha \Gamma (\frac{n + 1}{2})}
e^{-\frac{1}{2}{(q\theta)}^2}(q\theta )[3 - n + 2\frac{n - 1}{n + 1}{(q\theta)}^2],\ \ \ 
\end{equation}

\noindent For instance, in the case $n = 1$ one has the wavelet $\psi = (q\theta)
e^{-\frac{1}{2}{(q\theta)}^2}$, that is the optimal pseudo-filter to be used to detect 
a Gaussian signal on 1D spectra. In this case the detection level is 
(see equation(24)) ${\mathcal D}_w = AD^{-1/2}$.

\section {Exponential source on a background} \label{exponential}	
    
Typical example in astrophysics is the exponential disk associated to spiral galaxies. 
Another interesting application could be in some areas of physics where the profile 
expected for the signal associated to the detection of some particles
could be of exponential type.

Let us assume that the source and background can be represented by

\begin{equation}
s(x) = e^{-\frac{x}{\lambda}},\ \ \ P = Dq^{-\gamma}.
\end{equation}

In this case: 

\begin{equation}
\tau (q) = \beta {\lambda}^n{[1 + {(q\lambda )}^2]}^{-\frac{n + 1}{2}},
\end{equation}

\noindent where $\beta = {(\frac{2}{\pi})}^{1/2}, 1, 2{(\frac{2}{\pi})}^{1/2}$ for
$n = 1, 2, 3$, respectively, and equations (11) give

\begin{equation}
a = \frac{{\beta}^2{\lambda}^{n - \gamma}}{2D}\frac{\Gamma (m)
\Gamma (1 + \frac{n - \gamma}{2})}{\Gamma (n + 1)},
\end{equation}
\begin{equation}
 b = - ma,\ \ \  c = \frac{n + 1}{n + 2}m(1 + m)a,\ \ \ 
m\equiv \frac{n + \gamma}{2}
\end{equation}

\noindent and the pseudo-filter is

\begin{equation}
\tilde{\psi} (q) = \frac{2L}{\alpha \beta}
{(q\lambda)}^{\gamma}{[1 + {(q\lambda )}^2]}^{-\frac{n + 1}{2}}
[1 + \frac{\gamma - n}{2}(n + 1) +
M\frac{{(q\lambda)}^2}{1 + {(q\lambda)}^2}],\ \ \ 
\end{equation}

\begin{equation}
L\equiv \frac{\Gamma (n + 1)}
{\Gamma (\frac{n + \gamma}{2})\Gamma (2 + \frac{n - \gamma}{2})},\ \ \ 
M\equiv \frac{n - \gamma}{n + \gamma}(n + 1)(n + 2).
\end{equation}

In figure~\ref{fig2}  appear the optimal pseudo-filters for the 1D, 2D and 3D cases,
respectively, and power spectrum with indexes $\gamma = 0, 1, 2, 3$. 
The filter profiles are more extended than in
the Gaussian source, as one can expect from the more gentle
fall of the exponential source. 

An interesting case is $\gamma = n$, then the pseudo-filter is

\begin{equation}
\tilde{\psi} (q) = \frac{2n}{\alpha \beta}{(q\lambda )}^n
{[1 + {(q\lambda )}^2]}^{-\frac{n + 1}{2}}.
\end{equation}

a) Exponential source and white noise:

For this subcase $\gamma = 0$, then equation (36) leads to the pseudo-filter

\begin{equation}
\tilde{\psi} (q) = \frac{2}{\alpha \beta}\frac{\Gamma (n + 1)}
{\Gamma (\frac{n}{2})\Gamma (2 + \frac{n}{2})}
{[1 + {(q\lambda )}^2]}^{-\frac{n + 1}{2}}
[1 + \frac{n}{2}(n + 1) +
\frac{(n + 1)(n + 2){(q\lambda)}^2}{1 + {(q\lambda)}^2}].\ \ \ 
\end{equation}

For instance, for an exponential structure to be optimally detected in a 1D spectrum, 
we must use

\begin{equation}
\tilde{\psi} (q) = \frac{8}{{(2\pi)}^{1/2}}\frac{{(q\lambda)}^2}{{[1 + {(q\lambda )}^2]}^2}.
\end{equation}

b) Exponential source and $1/f$ noise:

An interesting case is $n = \gamma = 1$, equation (36) gives

\begin{equation}
\tilde{\psi} (q) = {(\frac{\pi}{2})}^{1/2}\frac{(q\lambda )}{1 + {(q\lambda )}^2}.
\end{equation}

\section {Simulations of one-dimensional Gaussian sources and $1/f$ noise} \label{simulations}

In order to test some of the ideas proposed in previous sections, we simulated the case of
one-dimensional Gaussian sources on a background. 
The kind of background simulated is the well-known $1/f$ noise.
This kind of noise appears very often in many devices
in experimental physics.
Further simulations 
with 2-dimensional
data and realistic realizations of noise will be carried on in future work.

For the sake of simplicity, all the simulated sources have the same amplitude
and that is set to be 1 (in arbitrary units). 100 of these
sources were deployed over a 32768 pixel 'field'. The number of sources and
the size of the field were selected in order to have enough sources for 
statistical studies, to avoid (as far as possible) the overlapping
of the sources and to minimize the
contribution of the sources to the total dispersion of the simulations.
 The width of the Gaussian profiles was chosen to be $\theta
\simeq 3\theta_{p}$: this is the case for a pixel of $1.5^{\prime}$ and 
a Gaussian with a FWHM of $5^{\prime}$. Noise was added so that the signal-to-noise
ratio of the sources, defined as the ratio $A/\sigma_{b}$ (where $A$ is 
the amplitude of the source and $\sigma_{b}$ is the standard deviation of
the noise), assumes values of $2,3,4$ and $5$. 
Finally, the optimal
filter, given by eq. (29) with $n=1$, was applied to the image.

To compare with a more traditional filtering scheme, we filtered the 
images also with a Gaussian of width equal to $\theta$ 
and a Mexican Hat wavelet of width equal to $\theta$. 
This is a rather naive usage of the Mexican Hat wavelet and the Gaussian
source because the optimal width for these filters in the general case 
is not the source
scale (Cay\'on 2000, Vielva 2000), but it serves us well because
what we intend is to compare how do filters work when we have
no further information about the data (i.e., the optimal scale, which is 
different for each background).
The result of these simulations is shown in tables~\ref{tb1} and
~\ref{tb2}.
Table~\ref{tb1} refers to the original simulations.
It shows the original signal-to-noise ratio of each simulation
as well as statistical quantities of interest such as the dispersion
of the map and the mean amplitude of the sources in it.
Finally, it shows the number of sources directly detected
from the simulations above $3\sigma$ and $5\sigma$ thresholds and the
number of spurious detections above these tresholds. As expected,
only a few sources are detected, except for the most favorable cases
(high original signal-to-noise rato and low detection threshold).
The small bias in the mean measured amplitude is due to
pixelization effects.  
Table~\ref{tb2}  refers to the simulations in table~\ref{tb1} after
filtering with 
a Gaussian of width $\theta$, a 
Mexican Hat wavelet of width $\theta$ and the optimal pseudo-filter. 
Each row in table~\ref{tb2} 
corresponds to the same row in table~\ref{tb1}.

A Gaussian filter  smoothes the image, removing small-scale noise. It also
smoothes the source peaks, thus lowering the amplitude of detected
sources. 
For the case of $1/f$ noise the dominant 
fluctuations  appear at large scales and are not affected by the Gaussian filter. 
The large-scale features may contribute to contamination in two different ways:
they can conceal sources in large 'valleys' and can produce spurious
peaks. None of these effects can be avoided with a Gaussian filter.
On the other hand, the smoothing effect of the Gaussian filter 
takes place normally and lowers the amplitude of the sources.
Therefore, the number of true detections is smaller than in the
non-filtered image, and the spurious detections are not removed even
in the highest $s/n$ case. The gains, indicated in column 6, clearly
reflects this situation ($g<1$).

The Mexican Hat wavelet has a better performance under
$1/f$ conditions. 
The
Mexican Hat removes large-scale fluctuations, allowing the 'hidden'
sources to arise above the detection threshold. For example, in the 
case of original $s/n = 2.95$ there were 47 sources above the $3\sigma$
level and only 1 above the $5\sigma$ level. 
After filtering with the Mexican Hat, there are 93 detections above
$3\sigma$ level and 64 above the $5\sigma$ level, a significant improvement.
The number of spurious sources remains almost untouched.

The optimal pseudo-filter  also deals with the
large-scale structure. It is constructed to enhance all
fluctuations in the source scale, while removing 
fluctuations that arise in other scales. In addition, 
it is required to be unbiased with respect to the amplitude.
In practice, the amplitude is slightly underestimated due
to the propagation of pixelization effects. This small
bias is lower than a $10\%$ and 
can be calibrated in any case. 
In the $1/f$ case the number of
true detections is higher than in the Mexican Hat case
and the number of spurious sources is comparable or
slightly reduced. Only in the case of low initial signal-to-noise ratio
the number of spurious detections is greater. This is due 
to the fact that this pseudo-filter enhances all
fluctuations in the source scale. Future work
will take care of this weakness of the method
including more information about the shape 
of the sources.
For the case of initial signal-to-noise ratio
of 2.95 we find 94 sources (of 100) and 8
spurious detections (a reliability close to $10\%$)
above the $3\sigma$ level, a result very
similar to the obtained with the Mexican Hat.
Above the $5\sigma$ detection level the optimal
pseudo-filter finds 79 sources where the Mexican Hat
found only 64. The number 
of spurious sources have not increased significantly (from
4 to 5).
For higher initial signal-to-noise ratios 
the completeness and reliability quickly improve. 

The gain obtained with the optimal pseudo-filter is
greater than the one obtained with the Mexican Hat.
It can be analytically calculated, using eqs. (6,17)
for the pseudo-filter and its equivalents for the 
Mexican Hat, leading to:

\begin{equation}
g_{op}/g_{mh} = \Big[ \frac{4}{\pi}\frac{\Gamma(\frac{1+\gamma}{2})\Gamma(\frac{5-\gamma}{2})}
{1+ \frac{ (1-\gamma) ^2}{2(1+\gamma)}}   \Big]^{1/2}
\end{equation}

\noindent for the one-dimensional case.
This formula holds while $\gamma \leqslant 1$.
According to
equation (40), the ratio $ g_{op}/g_{mh}$ is 1.41 in the case $\gamma=0$ and 1.13 in the
case $\gamma=1$. The mean observed ratio in the simulations is 1.31 and 1.08 respectively,
and fits well with our expectations. As a conclusion we have that optimal filter
gives higher gains than the classical Mexican Hat filter.

In figure~\ref{fig4} 
\notetoeditor{Figure 3 is a landscape-styled figure.}
an example of the simulations is shown. In the top panel there is 
a 500 pixel wide 
subsection of 
the $s/n=3$, $\gamma=1$ simulation. This 
subsection corresponds to a region in which 
the large-scale noise has a positive value. Four sources
are present in this area, all of them arising above the $3\sigma$ 
level (indicated
with a dotted line). The position of the sources are marked with an asterisk in
the lower panel. Additionally, there are three peaks, corresponding to 
background fluctuations, that arise above the $3\sigma$ level.
The second panel from the top shows the image after filtering with
the optimal filter. The large-scale features have been removed and also
the small-scale noise is reduced. The sources have been amplified with respect
to the original map and now all of them reach the $3\sigma$ level but the
spurious peaks have been removed. The
amplitudes of the sources remain unbiased and close to the true value of 1.
In the third panel from the top there is the image after filtering
with a Gaussian. The whole image has been smoothed and now one of
the sources barely reaches the $3\sigma$ level. The large-scale fluctuations
remain untouched and all the spurious peaks remain in the filtered 
image. 
In the bottom panel we see the image after filtering with the Mexican
Hat. The large-scale fluctuations are also removed as well as the small-scale
noise, as in the case of the optimal pseudo-filter.
 Nonetheless, the gain is lower and only three of the sources reach
the $3\sigma$ level. Additionally, it is found that the small-scale noise removal
is less efficient in the case of the Mexican Hat.

\section {Extraction of sources} \label{extraction}

The optimal pseudo-filter gives the position and an unbiased estimator of the amplitude
of the source. We propose to make the extraction of the source on real space, i. e.
one subtracts the function $A\tau (|\vec{x} - \vec{x_o}|)$, being $\tau$ the given profile and
$A$ the estimated amplitude,
at the position of the source $\vec{x_o}$.

From the practical point of view, in order to select the appropriate sources (with a 
given scale and avoiding to select spurious detections if the background and/or noise
are manifest at scales comparable to the sources) we can operate with the optimal 
pseudo-filter at other different scales $R$ as given by equation (10) but with 
$\tilde{\psi} (qx), x\equiv R/R_o$. If the scale that gives the maximum
do not correspond to the scale we are looking for then 
this is a spurious source (or another type 
of source with a different scale). As an additional check, we can calculate the source profile
in the pseudo-filter space
nearby any real source, e. g. for a Gaussian profile the behaviour around the maximum $R_o\approx R_s$
must be

\begin{equation}
<w(R)> = Ax^{\gamma }{(\frac{2}{1 + x^2})}^m[1 + \frac{n - \gamma }{2}\frac{1 - x^2}{1 + x^2}],\ \ 
x\equiv \frac{R}{R_o},\ \ \ m\equiv \frac{n + \gamma }{2},
\end{equation}

\noindent and an analogous behaviour can be found for the exponential profile. If a detected source
do not follows such a behaviour then it would be consider as
a false detection and must be deleted from the initial
catalog.

\section {Conclusions} 	 \label{conclusions}   

We have introduced for the first time (as far as the authors know) the concept of
{\it optimal} pseudo-filter to detect/extract spherical sources on a background modelled by
a (homogeneous \& isotropic) random field characterized by its power spectrum.
We have obtained a generic analytical formula that allows to calculate such a pseudo-filter 
either in Fourier or real space as a function of the source profile and power spectrum 
of the background. The psesudofilter is an unbiased an efficient estimator
of the amplitude of the sources.

We have applied the previous formula to the cases of a Gaussian and an exponential profile
and studied scale-free spectra. In particular, we have remarked the interesting cases of 
white noise and $1/f$ noise. We have calculated the detection level for the physically 
interesting cases of spectra, images and volumes. For some particular cases, the optimal 
pseudo-filters are wavelets (e. g. a Gaussian source embedded in white noise in the 2D case).

We have simulated Gaussian sources embedded in a $1/f$ noise in order to see the
performance of the optimal filter against the Mexican Hat wavelet. In the last case the gain is lower,
the noise removal is less efficient and the number of real 
detections is smaller. We also remark that
filtering with a Gaussian window is not the optimal procedure.

The extraction of the sources identified at a certain scale is proposed to be done directly
on real space. At the location of the source $\vec{x_o}$ one subtracts the function 
$A\tau (|\vec{x} - \vec{x_o}|)$, being $\tau$ the given profile.

All the calculations assume that the overlapping of nearby sources is negligible and the 
contribution of the sources to the background is also negligible. This is a very good 
approximation in many cases of interest at least above a certain flux level.

We remark the advantages of our method over traditional ones (DAOfind, SExtractor): we do not
need to assume a smooth background and/or some filters (e.g. Gaussian)
in order to detect the sources. For some astrophysical cases
(CMB) the background can be complex so a smooth surface could be not a reasonable assumption.
However, we need to assume the profile of the source and statistical properties of the background
in order to get the optimal filter. The main advantage of our method is the amplification effect (gain)
going to pseudo-filter space.

Generalization of these studies, considering different kind of sources (including
non-centrally symmetric ones), are now being undertaken.
The applications of this type of methodology is without any doubt
of interest not only for Astrophysics/Cosmology but for other sciences.

\acknowledgments

This work has been supported by the Comision Conjunta Hispano-Norteamericana de 
Cooperaci\'on Cient\'\i fica y Tecnol\'ogica ref. 98138, Spanish DGESIC Project no. 
PB98-0531-c02-01, FEDER project no. 1FD97-1769-c04-01 and the EEC project INTAS-OPEN-97-1992 
for partial financial support. J. L. S. acknowledges partial financial support from Spanish  
MEC and thanks CfPA and Berkeley Astronomy Dept. hospitality during year 1999.
D. H. acknowledges a Spanish M.E.C. PhD. scholarship.

\clearpage

\figcaption[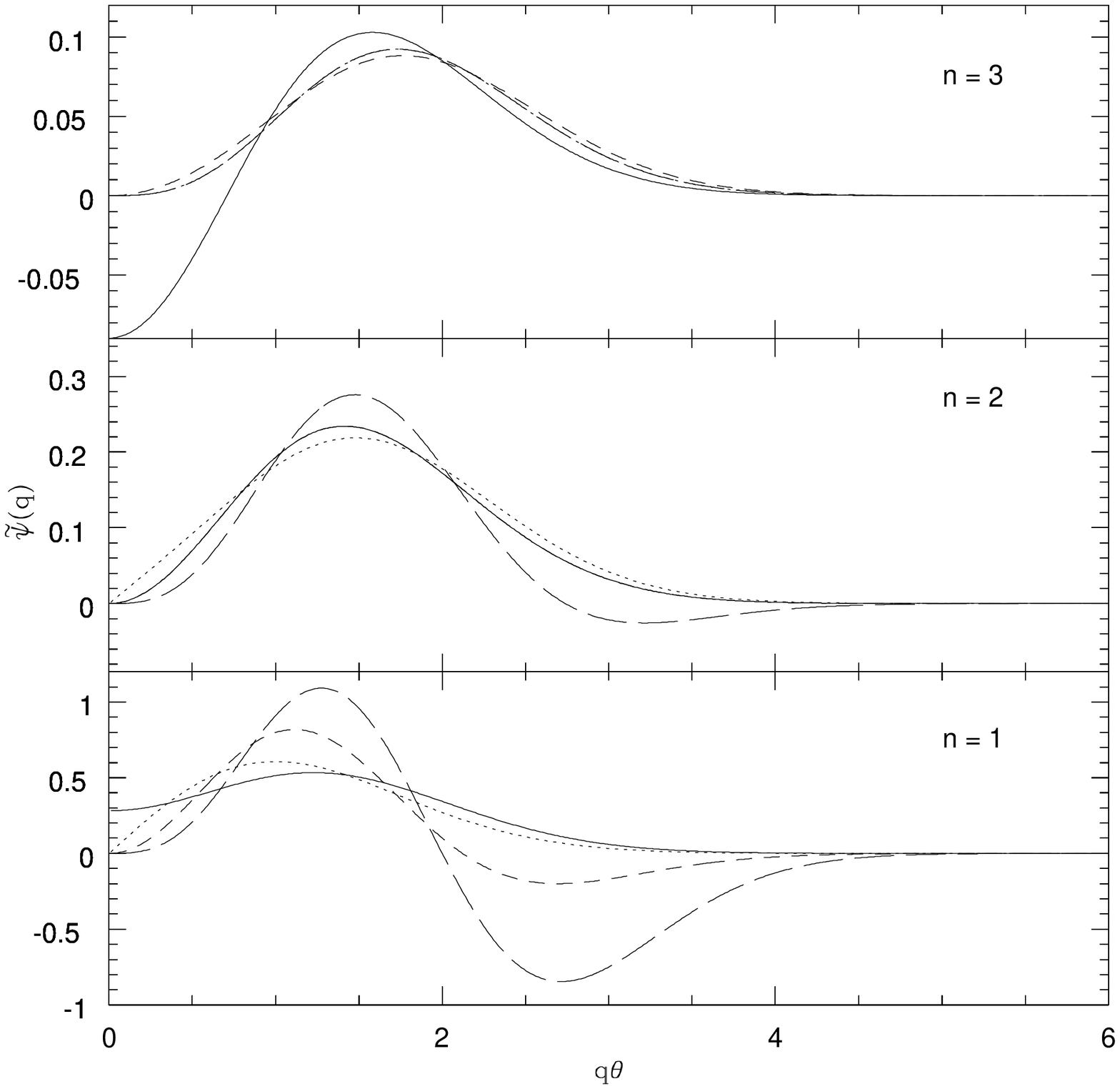]{Optimal filters for a Gaussian source in a background
$P(q) = Dq^{-\gamma}$ for $\gamma = 0$ (white noise, solid line), 
$\gamma=1$ ($1/f$ noise, dotted line), $\gamma=2$ (short-dashed line)
and $\gamma=3$ (large-dashed line). One, two and three-dimensional cases
are
represented. There is a degeneration in the case $n=3$, where
the $\gamma=1$ line overlaps with the $\gamma=3$ line, and in the case
$n=2$, where the $\gamma=0$ line overlaps with the $\gamma=2$ one. \label{fig1}}

\figcaption[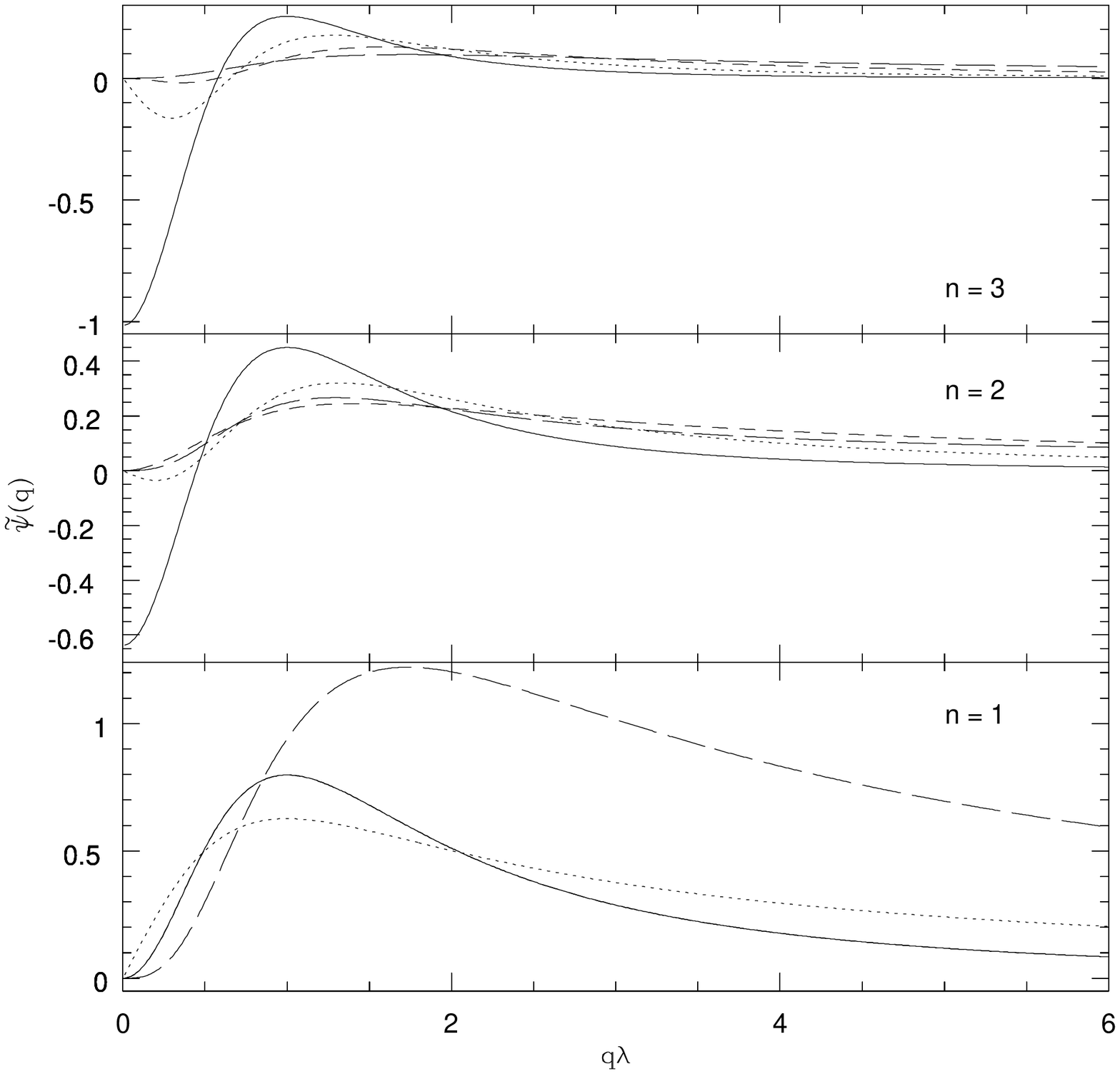]{Optimal filters for a exponential source in a background
$P(q) = Dq^{-\gamma}$ for $\gamma = 0$ (white noise, solid line), 
$\gamma=1$ ($1/f$ noise, dotted line), $\gamma=2$ (short-dashed line)
and $\gamma=3$ (large-dashed line). One, two and three-dimensional cases are
represented.  There is a degeneration in the case $n=3$, where
the $\gamma=0$ line overlaps with the $\gamma=2$ one.
\label{fig2}}

\figcaption[fig4.ps]{Performances of optimal pseudo-filter, 
Gaussian filter and Mexican Hat wavelet. In the top
panel a portion of a 32768-pixel one-dimensional simulation is shown.
In the simulation there are present Gaussian sources of width 
$\theta=3\theta_{p}$ and $1/f$ noise. The amplitude of the 
sources is  1 for all of them. The dispersion of the background is
set to be $1/3$. The second (third) 
panel from the top shows the map after filtering with
the optimal pseudo-filter (a Gaussian of width $\theta$). 
The bottom panel shows the map after filtering
with a Mexican Hat of width $\theta$. The positions of the sources are marked with asterisks.
In the three cases the $3\sigma$ level
of the resultant map is shown with an horizontal dotted line.
\label{fig4}}


\clearpage

 \begin{table}
 \begin{center}
 \begin{tabular}{ c c c c c c c c c c }
 \tableline
 \tableline
  $N$ & 
  $s/n$ & 
  $\sigma$ & 
  $\bar{A}$  & 
  $s/n_{m}$  & 
  $\sigma_{A_{op}}$ & 
  $d_{3\sigma} $  & 
  $e_{3\sigma}$ &  
  $d_{5\sigma} $  & 
  $e_{5\sigma}$ \\  
 \tableline
 1 &  2 &     0.5085 &     1.0670 &     2.0983 &     0.5075 &   19 &   12 &    0 &    0 \\ 
 2 &  3 &     0.3454 &     1.0195 &     2.9514 &     0.3432 &   47 &    9 &    1 &    0 \\ 
 3 &  4 &     0.2656 &     0.9994 &     3.7629 &     0.2582 &   72 &    6 &   10 &    0 \\ 
 4 &  5 &     0.2189 &     0.9890 &     4.5174 &     0.2070 &   90 &    9 &   36 &    1 \\ 
 \tableline
 \end{tabular}
\caption{\label{tb1}} 
 \tablecomments{Simulations.
The first column
shows the number of the simulation.
The second
column indicates the original signal-to-noise ratio, 
as explained in the text. 
Column 3 indicates the dispersion of the map. 
Column 4 shows the mean amplitude of the
sources as measured from the map and column 6 indicates the variance of the
source amplitudes. This quantity is not equal to the original signal-to-noise
ratio (column 2) because the sources contribute to the final dispersion
of the image, thus lowering the final, measured signal-to-noise ratio. 
Column 5 indicates the mean signal-to-noise ratio of the sources (that is,
the ratio between the quantities in columns 4 and 3). Columns 
7 and 8 indicate the number of sources detected at the $3\sigma$ level and
the number of $3\sigma$ detections that do not correspond to real
sources, respectively. Finally, columns 9 and 10 indicate the 
number of sources detected at the $5\sigma$ level and
the number of $5\sigma$ detections that do not correspond to real
sources, respectively.
 }
 \end{center}
 \end{table}

\clearpage

 \begin{table}
 \begin{center}
 \begin{tabular}{ c c c c c c c c c c }
 \tableline
 \tableline
  $N$ & 
  $\sigma_{f}$ & 
  $\bar{A}_{f}$  & 
  $s/n_{f}$  & 
  $\sigma_{A_{f}}$ & 
  $g$  & 
  $d_{3\sigma} $  & 
  $e_{3\sigma}$ &  
  $d_{5\sigma} $  & 
  $e_{5\sigma}$ \\  
 \tableline
 \multicolumn{10}{c}{GAUSSIAN FILTER} \\ 
 1 &     0.4816 &     0.7466 &     1.5504 &     0.4822 &     0.7389 &    8 &    3 &    0 &    0 \\ 
 2 &     0.3260 &     0.7237 &     2.2204 &     0.3216 &     0.7523 &   27 &    4 &    0 &    0 \\ 
 3 &     0.2494 &     0.7134 &     2.8602 &     0.2411 &     0.7601 &   46 &    3 &    0 &    0 \\ 
 4 &     0.2045 &     0.7076 &     3.4604 &     0.1929 &     0.7660 &   67 &    2 &    5 &    0 \\ 
 \multicolumn{10}{c}{MEXICAN HAT} \\ 
 1 &     0.2507 &     1.0070 &     4.0169 &     0.2475 &     1.9144 &   77 &   31 &   17 &    2 \\ 
 2 &     0.1798 &     0.9762 &     5.4298 &     0.1665 &     1.8397 &   93 &   16 &   64 &    4 \\ 
 3 &     0.1471 &     0.9642 &     6.5548 &     0.1250 &     1.7419 &   96 &    5 &   94 &    4 \\ 
 4 &     0.1292 &     0.9587 &     7.4214 &     0.1008 &     1.6429 &   98 &    2 &   98 &    2 \\ 
 \multicolumn{10}{c}{OPTIMAL PSEUDO-FILTER} \\ 
 1 &     0.2288 &     1.0159 &     4.4408 &     0.2217 &     2.1164 &   87 &   13 &   26 &    2 \\ 
 2 &     0.1673 &     0.9919 &     5.9305 &     0.1496 &     2.0094 &   94 &    8 &   79 &    5 \\ 
 3 &     0.1394 &     0.9825 &     7.0474 &     0.1131 &     1.8728 &   97 &    3 &   97 &    3 \\ 
 4 &     0.1244 &     0.9778 &     7.8592 &     0.0918 &     1.7398 &   99 &    1 &   99 &    1 \\ 
 \tableline
 \end{tabular}
\caption{\label{tb2}} 
 \tablecomments{Gaussian, Mexican Hat and optimal pseudo-filter results.
The first column
shows the number of the simulation,
the second shows the dispersion of the filtered map, the third column indicates
the mean estimated amplitude of the sources, the fourth column shows the 
mean signal-to-noise
ratio of the sources, column 5 indicates the dispersion of the measured amplitudes
of the sources, column 6 indicates the gain of the filter (calculated through
$(s/n)_{filter}/(s/n)_{original}$) and columns 7 and 8  
indicate the number of sources detected at a $3\sigma$ threshold and
the number of $3\sigma$ detections that do not correspond to real
sources, respectively.
 Finally, columns 9 and 10 indicate the 
number of sources detected at the $5\sigma$ level and
the number of $5\sigma$ detections that do not correspond to real
sources, respectively.}
 \end{center}
 \end{table}

\end{document}